\documentclass[aps,prb,preprint,floatfix,superscriptaddress,showpacs,amssymb,amsmath]{revtex4}
\usepackage{graphicx}

\begin{document}

\title{
Spin-F\"orster transfer in optically excited quantum dots
}
\author{Alexander O. Govorov}
\affiliation{Department of Physics and Astronomy, \\ Ohio
University, Athens, Ohio 45701-2979}
\date{\today } 

\begin{abstract}

The mechanisms of energy and spin transfer in quantum dot pairs
coupled via the Coulomb interaction are studied. Exciton transfer
can be resonant or phonon-assisted. In both cases, the transfer
rates strongly depend on the resonance conditions. The spin
selection rules in the transfer process come from the exchange and
spin-orbit interactions. The character of energy dissipation in
spin transfer is different than that in the traditional spin
currents. The spin-dependent photon cross-correlation functions
reflect the exciton transfer process. In addition, a mathematical
method to calculate F\"orster transfer in crystalline
nanostructures beyond the dipole-dipole approximation is
described.

\end{abstract}








\maketitle

A new field of research, spintronics, studies the principles of
manipulation of the spin degree of freedom in solids and molecules
\cite{book-spin}, whereas traditional electronics utilizes
electric charges. Spintronics is also closely related to quantum
information science since the spin is an important element of
quantum computing. In most cases, transport of spins in solids and
molecular systems comes from transfer or tunneling of charged
electrons and, therefore, is accompanied by electric currents. In
electronic materials, the electric interactions are often much
stronger than the spin-related ones and, therefore, when
manipulating charged particles with spins, usual electronics has
often obvious advantages compared to spintronics. However, spin
may have advantages over charge. In contrast to charge or mass,
the angular momentum or spin can be transferred without tunneling
or ballistic transport. One particular mechanism of spin transfer
without tunneling can occur in optically excited semiconductor
quantum dots (QDs); spin-polarized excitons can be transferred
between QDs via the long-range noncontact Coulomb interaction
\cite{Govorov-spin}. It is important to note that Coulomb
(F\"orster) transfer of spin in QDs becomes possible due to the
strong spin-dependent interactions in semiconductors, such as
spin-orbit and exchange couplings \cite{Govorov-spin}.

\begin{figure}[tbp]
\includegraphics*[width=0.6\linewidth, angle=90]{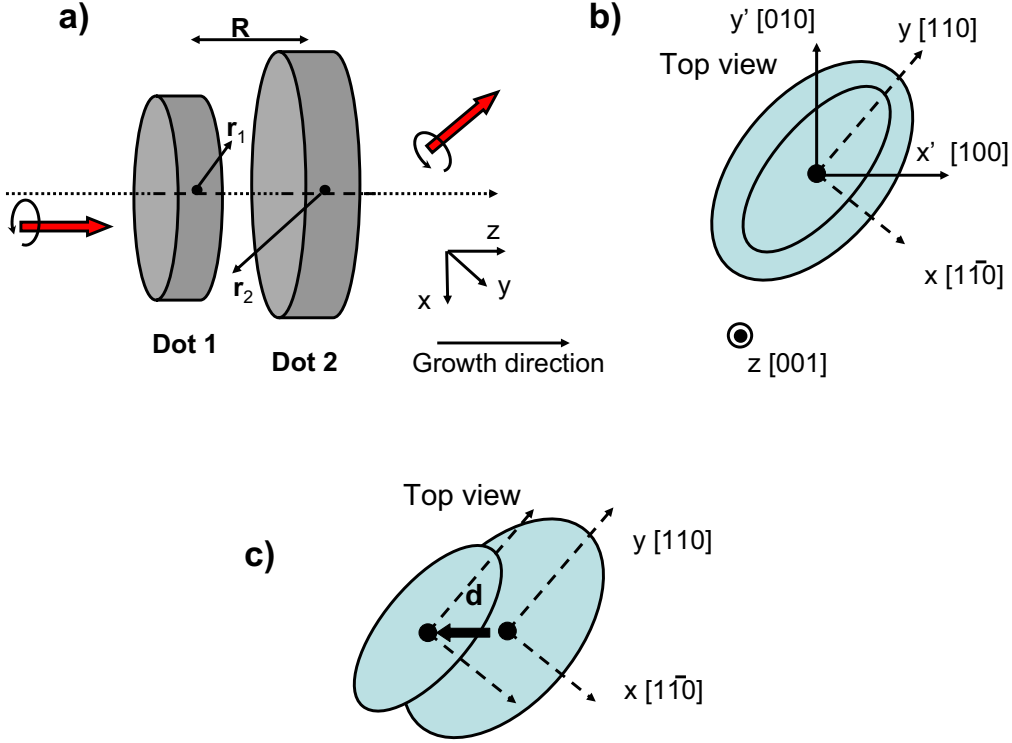}
\caption{(a) Schematics of the systems of two QDs. (b) Geometry of
a pair of self-assembled QDs and the corresponding
crystallographic axes. (c) Geometry of two QDs with broken
symmetry; the vector ${\bf d}$ describes the shift.} \label{fig1}
\end{figure}

\begin{figure}[tbp]
\includegraphics*[width=0.35\linewidth, angle=90]{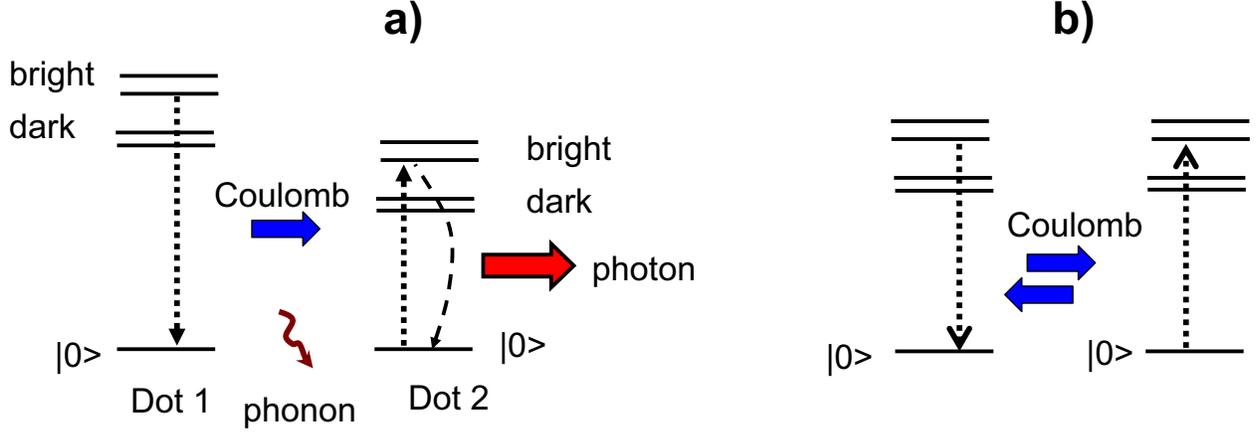}
\caption{Schematics of transfer processes between two QDs. (a)
Phonon-assisted mechanism. (b) Resonant coherent transfer.}
\label{fig2}
\end{figure}

Here we study theoretically the microscopic mechanisms of
spin-dependent F\"orster transfer in a molecule composed of two
self-assembled QDs (figs.~\ref{fig1},\ref{fig2}). In the typical
scheme of F\"orster transfer \cite{Forster}, an optically excited
exciton in QD1 ("donor") becomes transferred to QD2 ("acceptor")
via the Coulomb interaction (fig.~\ref{fig2}a). The traditional
methods to observe this transfer are time-resolved
photoluminescence (PL) spectroscopy \cite{Bawendi} and photon
correlations \cite{Petroff-preprint}. In the case of resonant
transfer in self-assembled QDs, the spin selection rules are
determined by the electron-hole exchange interaction in an exciton
and by the spin-orbit interaction in the valence band. In a
symmetric QD molecule, transfer occurs with conservation of the
exciton spin configuration, whereas, in QD molecules with broken
symmetry, the exciton spin becomes partially lost in the transfer
process. The transfer rates exhibit a strong dependence from the
exciton energy difference in a QD pair, $\Delta
E=E_{exc,dot1}-E_{exc,dot2}$. In the resonant regime $\Delta
E\approx0$, exciton and spin transfer is fast. In the nonresonant
regime, transfer can be assisted by acoustic phonons. Again, it
strongly depends on $\Delta E$. In contrast to the previous paper
on spin transfer \cite{Govorov-spin}, we include here the
electron-hole exchange interaction and consider the fine structure
of exciton states. In addition, we show that the dipole-dipole
approximation is not reliable for the typical inter-dot distances
in experimental structures and describe a method to compute
Coulomb matrix elements beyond the dipole approach. Our method is
valid when $R\gg a_{lattice}$, where $R$ is the inter-dot distance
and $a_{lattice}$ is the lattice period. In addition, we note that
the F\"orster transfer mechanism considered in this paper has the
electrostatic, near-field nature; this is in contrast to the
recent paper on the radiative coupling between QDs
\cite{Parascandolo}.

Experimentally, F\"orster transfer of excitons can be studied
using time-resolve photoluminescence  \cite{Bawendi} or the photon
cross-correlation method \cite{Petroff-preprint,TB,molecules}.
Experiments on energy transfer in nano-structures were done with
colloidal nano-crystals \cite{Bawendi} and recently with
self-assembled $InAs$ QDs \cite{Petroff-preprint}. It is important
to note that QDs can also be coupled via tunneling
\cite{Manfred-Gerhard}. However, the tunneling amplitude decreases
exponentially with increasing the distance between QDs and with
the hight of the barrier between QDs. At the same time, the
F\"orster transfer rate demonstrates a power law ($\propto
R^{-6}$) and is independent of the hight of the potential barrier
between QDs. To avoid tunneling in a QD molecule, one can grow
$AlGaAs$ barrier between the QDs or one can use QDs with stronger
confinement. Another important factor is that the resonance
conditions for F\"orster and tunneling transfers are different.
Therefore, it seems to be possible, by using self-organization
growth, to design self-assembled QDs with predominantly F\"orster
transfer. The recent experiment \cite{Petroff-preprint} indicates
that indeed such pairs of QDs with Coulomb-induced coupling can be
designed using the self-organization growth method. Another recent
experimental paper (ref.~\cite{bridgedQDs1}) describes the
spin-response of colloidal QDs bridged with bio-molecules which
may assist direct tunnel transport between nano-crystals
\cite{bridgedQDs1,bridgedQDs2}.

The spin current of mobile polarized electrons is always
accompanied by the electric charge flow.  This brings back the old
issue of energy dissipation in electronic devices, given by the
Joule heat, $Q={\bf j}{\bf E}$ (here ${\bf j}$ is the electric
current  and  ${\bf E}$ the driving electric field). In the case
of Coulomb-induced transfer between QDs (figs.~\ref{fig2}a and
\ref{fig9}), the energy dissipation has a different character and
comes from phonon-assisted relaxation. The energy loss in this
process is equal to the energy level difference in the donor and
acceptor QDs: $\Delta
E=E_{exc,dot1}-E_{exc,dot2}=\hbar\omega_{ph}$. In such processes,
the energy $\Delta E$ turns into the phonon energy
$\hbar\omega_{ph}$. If a pair of QDs is resonant ($\Delta
E\sim0$), the coupling between QDs can become coherent
\cite{coherent,coherent2,coherent3}; such a process of spin
transfer resembles Rabi oscillations between QDs
(fig.~\ref{fig2}b). Coherent spin transfer can occur without
dissipation.

The paper is organized as follows: the section~1 describes the
model of the QD system, the section~2 includes the results on
spin-dependent transfer rates, the section~3 is devoted to the
phonon-assisted transfer, and the sections~4 and 5 are about the
photon correlation functions and transfer under strongly resonant
conditions.

\section{ Model }

We now consider a model of a pair of self-assembled QDs
\cite{Petroff-Bimberg,Manfred,Petroff-preprint}. Our model
incorporates two oblate asymmetric QDs (figs.~\ref{fig1}a and b);
the vertical, $z$-size of QDs is assumed to be much less than the
lateral ones. To model the lateral motion of electrons and holes,
we use the harmonic functions \cite{Govorov-PRB} with the
characteristic lengthes, $l_{e(h),x}=\sqrt{\hbar/\omega_x^{e(h)}
m_{e(h)}}$ and $l_{y}^{e(h)}=\sqrt{\hbar/\omega_{y}^{e(h)}
m_{e(h)}}$, where $\omega_{x(y)}^{e(h)}$ are the
harmonic-oscillator frequencies for electrons (e) and holes (h),
$m_{e(h)}$ are the effective masses of particles, and $(x,y)$ are
the in-plane coordinates. The complete envelope functions used
below have a form:
$\phi_{e,k}=A_{e,k}e^{-x^2/2l_{e,x}^2-y^2/2l_{e,y}^2}sin[\pi(z-z_{k})/L]$
and
$\phi_{h,k}=A_{h,k}e^{-x^2/2l_{h,x}^2-y^2/2l_{h,y}^2}sin[\pi(z-z_{k})/L]$,
where $i=x,y$, $k=1,2$ is the dot number, $z_{k}$ is the
$z$-coordinate of the center of $k$-dot, $A_{k,e(h)}$ are the
normalizing coefficients, and $L$ is the "vertical" size of QDs.
In the following, we will use the system of coordinates $(x,y)$
(fig.~\ref{fig1}b) which corresponds to the typical orientation of
elongated self-assembled QDs grown on the $[001]$ surface
\cite{exchange}. The lowest excitonic states in QDs responsible
for PL are composed of heavy holes and electrons and correspond to
the $s$-like envelope wave function. Taking into account the only
heavy-hole wave functions, we can write the spin Hamiltonian of
exciton of an individual QD in the following form
\cite{exchange,bookIP}:

\begin{eqnarray}
H_{e-h}^{spin}=a_z\hat{j}_z\hat{s}_z+\sum_{i=x,y,z}b_i\hat{j}_i^3\hat{s}_i,
\label{Hexc}
\end{eqnarray}
where $\hat{s}_i$ is the electron spin matrixes and $\hat{s}_i$
are the 2x2 angular-momentum operators of heavy holes. The
exchange parameters $a_z$ and $b_i$ depend on a particular QD. By
using the operator (\ref{Hexc}), we find the exciton wave
functions and their energies. The bright excitons are composed of
the states with $J_{tot}=j_z+s_z=\pm1$:

\begin{eqnarray}
\psi_{x}^b=\frac{|+\frac{1}{2};-\frac{3}{2}>+|-\frac{1}{2};+\frac{3}{2}>}{\sqrt{2}},
\nonumber
\\
\psi_{y}^b=\frac{|+\frac{1}{2};-\frac{3}{2}>-|-\frac{1}{2};+\frac{3}{2}>}{\sqrt{2}},
\label{WaveFunctionsExciton1},
\end{eqnarray}
where we used the notation $|s_z;j_z>$. The corresponding energies
$\epsilon_{x}^b=-(\frac{3}{4}a_z+\frac{27}{16}b_z)+\frac{3}{8}(b_x-b_y)$
and
$\epsilon_{y}^b=-(\frac{3}{4}a_z+\frac{27}{16}b_z)-\frac{3}{8}(b_x-b_y)$.
The lower indexes $x,y$ reflect the character of spin orientation
in an exciton and the optical selection rules. In the PL process,
the excitons $\psi_{x}^b$ and $\psi_{y}^b$ create photons with
linear $x$ and $y$ polarizations, respectively. The dark excitons
are composed of the states with $J_{tot}=j_z+s_z=\pm2$:

\begin{eqnarray}
\psi_{x}^d=\frac{|+\frac{1}{2};+\frac{3}{2}>+|-\frac{1}{2};-\frac{3}{2}>}{\sqrt{2}},
\nonumber
\\
\psi_{y}^d=\frac{|+\frac{1}{2};+\frac{3}{2}>-|-\frac{1}{2};-\frac{3}{2}>}{\sqrt{2}}.
\label{WaveFunctionsExciton2}
\end{eqnarray}
Their energies
$\epsilon_{x}^b=-(\frac{3}{4}a_z+\frac{27}{16}b_z)+\frac{3}{8}(b_x+b_y)$
and
$\epsilon_{y}^b=-(\frac{3}{4}a_z+\frac{27}{16}b_z)-\frac{3}{8}(b_x+b_y)$.
Typically, the two lowest states in the exciton spectrum are dark
whereas the two upper ones are bright (fig.~\ref{fig2}). In this
model, the normal magnetic field does not lead to mixing between
dark and bright states, inducing an additional splitting in the
pairs of states. In the limit $B\rightarrow\infty$, the wave
functions become the states in which the angular momentum is a
good quantum number.

The exchange spin-dependent interaction in excitons and the
dark-bright energy splitting are quite strong, about $0.5~meV$
\cite{exchange}. The other types of interaction in crystalline
QDs, inter-dot Coulomb and electron-phonon couplings, can be
weaker and we are going to involve them as perturbation. We note
that the intra-dot Coulomb interaction is quite strong, but it
does not lead to the inter-dot exciton transfer; it mostly shifts
down the exciton energies. Then, the perturbation Hamiltonian is

\begin{eqnarray}
\hat{H}_{perturb}=U_{Coul}(r_1,r_2)+\hat{H}_{e-ph}+\hat{H}_{h-ph},
\label{Hperturb}
\end{eqnarray}
where $U_{Coul}$ is the inter-dot Coulomb interaction. The
operators $\hat{H}_{e-ph}$ and $\hat{H}_{h-ph}$ represent the
interaction between acoustic phonons and particles.

\section{ Spin-dependent Coulomb matrix elements}

First we compute the inter-dot Coulomb matrix elements. The
complete set of electron-hole wave functions includes 8 states:

\begin{eqnarray}
|\pm\frac{1}{2},\pm\frac{3}{2};k> , \hskip 0.4 cm
|\pm\frac{1}{2},\pm\frac{3}{2};k>, \label{set2}
\end{eqnarray}
where $k=1,2$ in the QD index. For the one-exciton states, we have
$|s_z',j_z';1>|0,2>$ and $|0;1>|s_z',j_z';2>$; here $|0;k>$
denotes the state of the $k$-dot without an exciton. Then we write
the inter-dot Coulomb matrix elements as:

\begin{eqnarray}
<0;1|<s_2,j_2;2|U_{Coul}|s_1,j_1;1>|0,2>. \label{CoulombMatEl}
\end{eqnarray}
In most papers, the matrix elements (\ref{CoulombMatEl}) are
calculated within the dipole-dipole approximation which is valid
in the limit $R\ll l_{dot}$, where $R$ is the inter-dot distance
and $l_{dot}$ is a characteristic size of QDs ($l_{dot}\sim
l_{e(h),x(y)}$). Now we are going to use a method beyond the
dipole-dipole approximation. Namely, we are going to use the
quantity $a_{lattice}/R$ as a small parameter, where $a_{lattice}$
is the crystal-lattice period. Since, $a_{lattice}\ll l_{dot}$,
our approximation is much better compared with the standard
dipole-dipole approximation, $R\gg l_{dot}$. To evaluate the
matrix element (\ref{CoulombMatEl}), it is convenient to return to
the pure electron representation and to consider two electrons in
each QD explicitly:

\begin{figure}[tbp]
\includegraphics*[width=0.7\linewidth, angle=90]{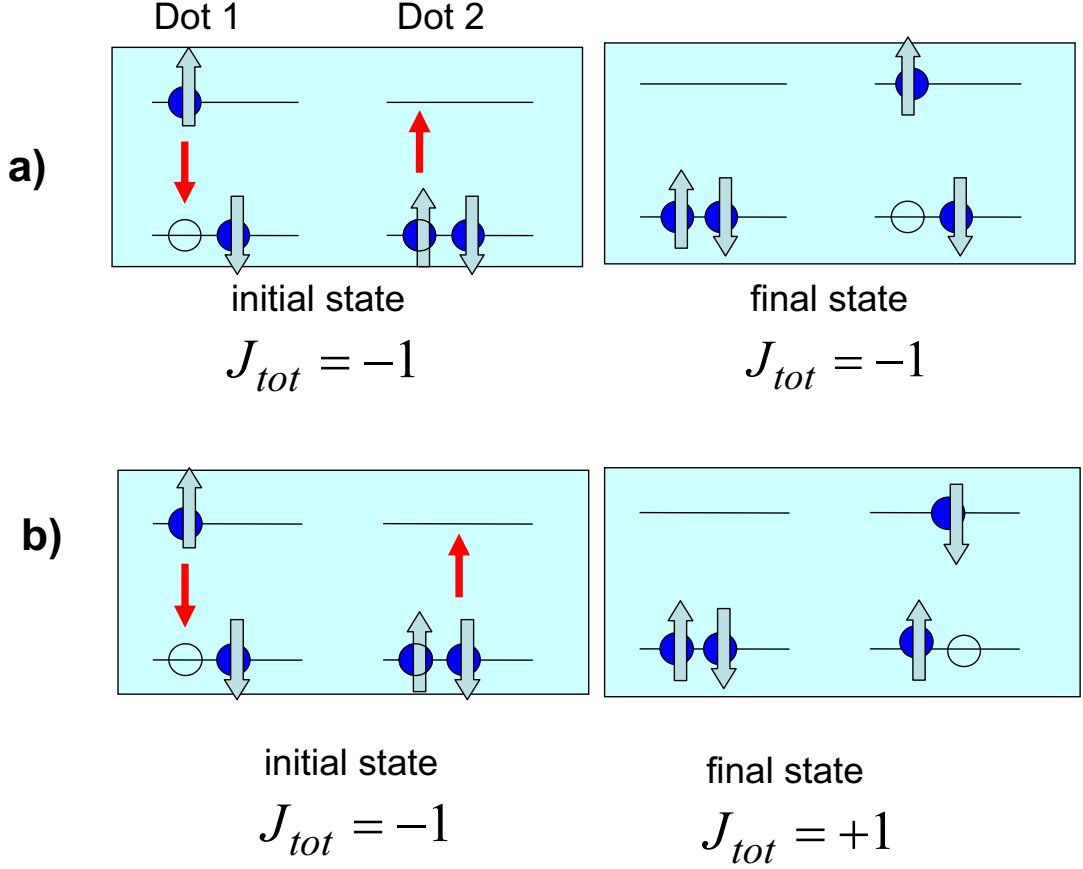}
\caption{ Electron configurations for the initial and final states
of the transfer processes with  (a) and without (b) conservation
of the exciton angular momentum. The process (b) becomes possible
in QD pairs with broken symmetry ($d\neq0$).} \label{fig3}
\end{figure}

\begin{eqnarray}
\nonumber
M_{|s_1,j_1;dot1>\rightarrow |s_2,j_2;dot2>}=<0;1|<s_{2},j_{2};2|U_{Coul}|s_{1},j_{1};1>|0,2>= \\
\int d\xi_1 d\xi_1' d\xi_2 d\xi_2'
[\Psi_{j_{1},1}(\xi_1)\Psi_{-j_{1},1}(\xi_1')\Psi_{s_2,2}(\xi_2)\Psi_{j_2,2}(\xi_2')]^*
\nonumber \\ U_{Coul}(r_1, r_1', r_2, r _2')
\Psi_{j_{1},1}(\xi_1)\Psi_{s_{1},1}(\xi_1')\Psi_{j_{2},2}(\xi_2)\Psi_{-j_{2},2}(\xi_2'),
\nonumber \\ \label{CoulombMatElem2}
\end{eqnarray}
here $\xi_1=({\bf r}_1,\sigma_1)$ and $\xi_1'=({\bf
r}_1',\sigma_1')$ are the spatial and spin coordinates of two
electrons in QD1; $\xi_2=({\bf r}_2,\sigma_2)$ and $\xi_2'=({\bf
r}_2',\sigma_2')$ are similar coordinates for QD2. $s_{1(2)}$ are
the z-components of spin in the conduction band of QD1(2);
$j_{1(2)}$ are the z-components of angular momentum of the valence
band-electrons. $\Psi_{j_k,k}$ is the electron state in the
valence band of the k-QD ($k=1,2, \ j_k=\pm3/2$). The function
$\Psi_{s_k,k}$ is the state for the conduction band
($s_k=\pm1/2$). Within the envelope-function approach
$\Psi_{j,k}(\xi)=\chi_{j}(r,\sigma)\phi_{h,i}(r)$ and
$\Psi_{s,k}(\xi)=\chi_{s}(r,\sigma)\phi_{e,i}(r)$, where
$\chi_{j}$ and $\chi_{s}$ are the Bloch functions in the valence
band (heavy-hole) and in the conduction band, respectively;
$\phi_{h(e)}(r)$ is the hole (electron) envelope function. The
F\"orster process involves one electron in QD1 with coordinate
$r_1$ and one electron in QD2 with coordinate $r_2$
(fig.~\ref{fig3}) and therefore we can integrate over $r_{1(2)}'$.
For smooth envelope functions and long-range Coulomb potential, we
can rewrite the integral (\ref{CoulombMatElem2}) as follows:

\begin{eqnarray}
\nonumber \sum_{\alpha_1,\alpha_2}
\phi_{e,1}(R_{\alpha_1})\phi_{h,1}(R_{\alpha_1})\phi_{e,2}
(R_{\alpha_2})\phi_{h,2}(R_{\alpha_2})\\ \nonumber
*\int_{\Omega_{\alpha_1}}\int_{\Omega_{\alpha_2}}d\Delta\xi_1d\Delta\xi_2
[\chi_{j_1}(\xi_1)^*\chi_{s_2}(\xi_2)^*U_{Coul}(R_{\alpha_1}+\Delta
R_{\alpha_1}, R_{\alpha_2}+\Delta
R_{\alpha_2})\chi_{s_1}(\xi_1)\chi_{j_2}(\xi_2)],
\\  \label{CoulombMatElem3}
\end{eqnarray}
where the summation is performed over all the unit cells in both
QDs; $\alpha_{1(2)}$ are the unit cell indexes;
$\Omega_{\alpha_k}$ and $R_{\alpha_k}$ are the unit cell volumes
and unit cell coordinates, respectively ($k=1,2$).
$\Delta\xi_k=(\Delta R_{\alpha_k},\sigma_k)$, where $\Delta
R_{\alpha_k}$ is the spatial coordinate relative to the center of
the $\alpha$-cell and $\sigma_k$ is the spin coordinate. Assuming
$a_{lattice}/l_{dot}\sim a_{lattice}/R\ll 1$, we expand the
Coulomb potential in terms of $\Delta R_{\alpha_k}$ and take into
account the leading term responsible for the F\"orster transfer:

\begin{eqnarray}
U_{Coul}(R_{\alpha_1}+\Delta R_{\alpha_1}, R_{\alpha_2}+\Delta
R_{\alpha_2})=\frac{e^2}{\epsilon} \frac{\Delta R_{\alpha_1}\Delta
R_{\alpha_2} - 3[(\Delta R_{\alpha_1} R_{\alpha_1,\alpha_2})
(\Delta
R_{\alpha_2}R_{\alpha_1,\alpha_2})]/|R_{\alpha_1,\alpha_2}|^2}
{|R_{\alpha_1,\alpha_2}|^3}, \nonumber \\ \label{CoulombPot}
\end{eqnarray}
where $R_{\alpha_1,\alpha_2}=R_{\alpha_1}-R_{\alpha_2}$. The
Coulomb potential was taken in the usual form:
$U_{Coul}=e^2/(\epsilon|r_1-r_2|)$. Changing the summation in
eq.~(\ref{CoulombMatElem3}) to integration, we obtain

\begin{eqnarray}
M_{|s_1,j_1;dot1>\rightarrow |s_2,j_2;dot2>}=
\frac{e^2}{\epsilon}\int
dR_1dR_2\phi_{e,1}(R_1)\phi_{h,1}(R_1)\phi_{e,2}
(R_2)\phi_{h,2}(R_2) \nonumber \\
\int_{\Omega_1}\int_{\Omega_2}d\Delta\xi_1d\Delta\xi_2
[\chi_{j_1}(\Delta\xi_1)^*\chi_{s_2}(\Delta\xi_2)^* \frac{\Delta
R_1 \Delta R_2 - 3[(\Delta R_1 R_{12}) (\Delta R_2
R_{12})]/|R_{12}|^2}
{|R_{12}|^3}\chi_{s_1}(\Delta\xi_1)\chi_{j_2}(\Delta\xi_2)], \nonumber \\
\label{CoulombMatElem5}
\end{eqnarray}
where $R_{12}=R_1-R_2$ and $\Delta\xi_k=(\Delta R_k,\sigma_k)$. In
the next step, we derive the matrix elements for the heavy-hole
excitons using the Bloch functions, $\chi_{1/2}=|S>\uparrow$, \
$\chi_{-1/2}=|S>\downarrow$, \
$\chi_{+3/2}=(|X>+i|Y>)\uparrow/\sqrt{2}$, \
$\chi_{-3/2}=(|X>-i|Y>)\downarrow/\sqrt{2}$,

\begin{eqnarray}
\nonumber
M_0=M_{|s_1=+1/2,j_1=-3/2;dot1>\rightarrow|s_2=+1/2,j_2=-3/2;dot2>} \\
=M_{|s_1=-1/2,j_1=+3/2;dot1>\rightarrow|s_2=-1/2,j_2=+3/2;dot2>}
\nonumber
\\ =E_0 \int
dR_1dR_2F_{cv}(R_1,R_2)\frac{1-\frac{3}{2}\frac{(X_1-X_2)^2+(Y_1-Y_2)^2}{|R_1-R_2|^2}}{|R_1-R_2|^3}
\nonumber \\ \nonumber
M_1=M_{|s_1=+1/2,j_1=-3/2;dot1>\rightarrow|s_2=-1/2,j_2=+3/2;dot2>} \\
\nonumber
=M_{|s_1=-1/2,j_1=+3/2;dot1>\rightarrow|s_2=+1/2,j_2=-3/2;dot2>}^* \\
=-E_0 \int
dR_1dR_2F_{cv}(R_1,R_2)\frac{3}{2}\frac{[X_1-X_2-i(Y_1-Y_2)]^2}{|R_1-R_2|^5}.
\label{ME0}
\end{eqnarray}
Here $E_0=e^2d_0^2/\epsilon$,
$F_{cv}(R_1,R_2)=\phi_{e,1}(R_1)\phi_{h,1}(R_1)\phi_{e,2}
(R_2)\phi_{h,2}(R_2)$ and $d_0=<S|x|X> $. The matrix element
$<S|x|X>$ can also be written as $-\hbar/(i m_0)(P_{cv}/E_g)$,
where $P_{cv}$ and $E_g$ are the inter-band optical matrix element
and the band gap energy of the bulk crystal, respectively.

\begin{figure}[tbp]
\includegraphics*[width=0.7\linewidth, angle=90]{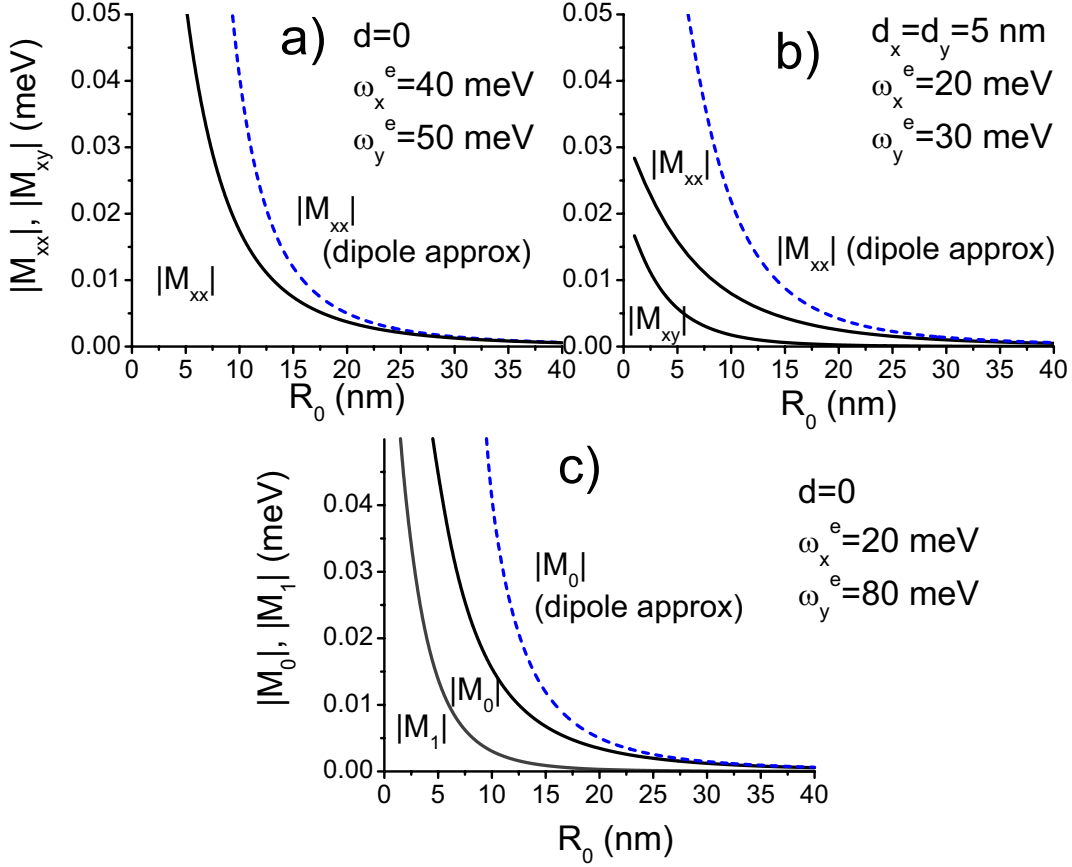}
\caption{ Calculated transfer amplitudes for QD pairs of various
parameters and symmetry. a) Transfer amplitudes for $d=0$.
$M_{xx}=M_{yy}$ describe the processes with spin conservation;
$|M_{xy}|=|M_{yx}|=0$. (b) Transfer amplitudes at $d\neq0$. (c)
Calculated transfer amplitudes $M_0$ (spin conservation) and $M_1$
(spin-flip) describing transfer processes at high magnetic fields
and $d=0$. In (c), the QDs are strongly asymmetric. } \label{fig4}
\end{figure}

The matrix element $M_0$ describes the transfer process with
conservation of spin, whereas $M_1$ relates to the spin-flip
process. The exciton states with $J_{tot}=\pm2$ have no matrix
elements in our model \cite{commentHL}. The transfer processes
with bright excitons (eq.~\ref{WaveFunctionsExciton1}) have the
following amplitudes:

\begin{eqnarray}
\nonumber M_{\psi^b_x,dot1\rightarrow\psi^b_x,dot2}=M_0+Re[M_1] \\
\nonumber M_{\psi^b_y,dot1\rightarrow\psi^b_y,dot2}=M_0-Re[M_1] \\
\nonumber M_{\psi^b_x,dot1\rightarrow\psi^b_y,dot2}=-i
Im[M_1] \\
M_{\psi^b_y,dot1\rightarrow\psi^b_x,dot2}=i Im[M_1]. \label{ME1}
\end{eqnarray}
The matrix elements (\ref{ME1}) strongly depend on symmetry.
Especially, it is related to the off-diagonal transfer processes
$x\leftrightarrow y$. The off-diagonal amplitudes can be written
as

\begin{eqnarray}
M_{\psi^b_x,dot1\rightarrow\psi^b_y,dot2}=M_{\psi^b_y,dot1\rightarrow\psi^b_x,dot2}^*=-E_0
\int
dR_1dR_2F_{cv}(R_1,R_2)\frac{3(X_1-X_2)(Y_1-Y_2)}{|R_1-R_2|^5}.
\label{ME2}
\end{eqnarray}
If the double-dot system is symmetric with respect to the
inversion operations $R_{1(2)}\rightarrow -R_{1(2)}$, $M_1=0$ and
the transfer process conserves the linear polarization of
excitons, i.e.  $x$-exciton in QD1 turns into $x$-exciton in QD2
and the same rule is applied to $y$-excitons. Therefore spin
information can be transferred without losses in the system with
spatial-inversion symmetry $R\rightarrow -R$. If the dots are
shifted with respect to each other in the $xy$-plane
(fig.~\ref{fig1}c), spin information in the transfer process
becomes partially lost since $M_1\neq0$; if the shift $d$ is
small, $M_1\propto d_xd_y$. The calculated amplitudes of various
transfer processes are shown in fig.~\ref{fig4}. For both QDs, we
used the following parameters: $L=2~nm$,
$\omega^h_{x(y)}=\omega^e_{x(y)}/3$, $m_e=0.07 m_0$, and $m_h=0.25
m_0$. The above parameters are typical for $InAs$-based QDs
\cite{Govorov-PRB}.

We note that it is very important to compute the matrix elements
beyond the dipole-dipole approach since the amplitudes
$x\leftrightarrow y$ vanish within the dipole-dipole
approximation. Also, the generalized dipole approach ($R\gg
a_{lattice}$) used in this paper is necessary to obtain reliable
numbers for all the matrix elements at inter-dot distances $R\sim
l_{dot}$ which are typical for experiments. The amplitudes of
processes with spin-flip, $|\psi^b_x,dot1> \rightarrow
|\psi^b_y,dot2>$ and $|\psi^b_y,dot1>\rightarrow|\psi^b_x,dot2>$,
cannot be obtained within the dipole-dipole approach:
$Im[M_1]\propto R^{-5}$ for $R\rightarrow\infty$. At the same
time, the F\"orster transfer elements with conservation of exciton
spin has the usual asymptotic behavior at $R\rightarrow\infty$:
$M_0\propto R^{-3}$.

In a normal magnetic field, the Hamiltonian (\ref{Hexc}) has an
additional term

\begin{eqnarray}
H_{e-h}^{mag}=\mu_B(g_e\hat{s}_z+\frac{g_{h,z}}{3}\hat{j}_z)B,
\label{HB}
\end{eqnarray}
where $B$ is the normal magnetic field, and $g_e$ and $g_{h,z}$
are the $g$ factors. The eigenstates of the Hamiltonian in a
strong magnetic field are pure states of the total angular
momentum: $|s_k=\pm1/2,j_k=\pm3/2>$. Thus, the transfer matrix
elements in the limit $B\rightarrow\infty$ are given by
eqs.~\ref{ME0}. The transfer process with conservation of spin is
given by the element $M_0$, whereas the spin-flip transfer
processes are given be $M_1$. Again, it is important to stress the
role of symmetry for exciton transfer with spin-flip. If the
double-dot system is cylindrically symmetric, the spin-flip
transfer processes vanish, $M_1=0$. In the case of asymmetric QDs,
$M_1\neq0$. Figure~\ref{fig4} shows the calculated amplitudes for
the case $d=0$ and strongly asymmetric QDs. One can see that the
spin-flip processes become important at small inter-dot distances.
Again, the dipole approach would not describe such spin-flip
effects.

\section{ Phonon-assisted Coulomb transfer}

\begin{figure}[tbp]
\includegraphics*[width=0.3\linewidth, angle=90]{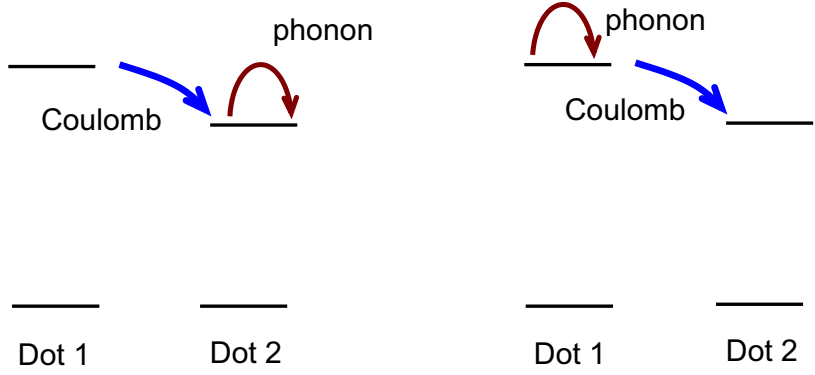}
\caption{ Two contributions to phonon-assisted transfer with
different virtual intermediate states.} \label{fig5}
\end{figure}

In real QD systems, it is very difficult to find QDs with the same
exciton energy. Therefore, one should involve acoustic phonons to
satisfy the energy conservation requirement (fig.~\ref{fig2}a).
The operator of exciton-phonon interaction reads

\begin{eqnarray}
\hat{H}_{exc-ph}=\sum_q\sqrt{\frac{\hbar q}{2\rho V
c_{ph}}}[\sigma_ee^{i{\bf q}{\bf
r}_e}(\hat{c}_q+\hat{c}_{-q}^+)+\sigma_he^{i{\bf q}{\bf
r}_h}(\hat{c}_q+\hat{c}_{-q}^+)] \label{Hph}
\end{eqnarray}
\cite{Maham}. Here $\hat{c}_q$ is the phonon annihilation
operator, $r_{e(h)}$ are the electron (hole) coordinates,
$c_{ph}=5.6*10^5~cm/s$ is the speed of longitudinal sound,
$\sigma_{e(h)}=-8.0~eV (1.0~eV)$ are the deformational potentials,
and $\rho=5.3~g/cm^3$ is the crystal mass density. The rate of
phonon-assisted transfer includes two second-order processes
(fig.~\ref{fig5}):

\begin{eqnarray}
W^{ph}_{dot1,\alpha\rightarrow
dot2,\alpha'}=\frac{2\pi}{\hbar}\sum_q|\frac{<dot2|\hat{H}_{exc-ph}|dot2><dot2|U_{Coul}|dot1>}{\Delta
E} \nonumber \\
+\frac{<dot2|U_{Coul}|dot1><dot1|\hat{H}_{exc-ph}|dot1>}{-\hbar
c_{ph}|q|}|^2\delta(\Delta E-\hbar c_{ph}|q|), \label{Wph}
\end{eqnarray}
where $|dot1>=|\alpha;1>|0;2>$ and $|dot2>=|0;1>|\alpha';2>$
denote the states in which an exciton is in QD1 or in QD2,
respectively; $\alpha=x,y$ is the spin index of exciton. The
notation $|0;k>$ means an empty QD.

Then, the rate (\ref{Wph}) is reduced to

\begin{eqnarray}
W^{ph}_{dot1,\alpha\rightarrow
dot2,\alpha'}=\frac{1}{2\pi\hbar}\frac{|M_{dot1,\alpha\rightarrow
dot2,\alpha'}|^2}{\Delta E^3}\frac{\hbar q_0^4}{\rho
c_{ph}}F(q_0)(N(\Delta E)+1), \label{Wph2}
\end{eqnarray}
where $M_{dot1,\alpha\rightarrow dot2,\alpha'}$ is the inter-dot
Coulomb matrix element between the excitonic states $\alpha$ and
$\alpha'$ given by eqs.~\ref{ME1} and \ref{ME2}, the index
$\alpha=x,y$, $q_0=\Delta E/(\hbar c_{ph})$, $F(q_0)$ is a
function given by an integral, and $N(\Delta E)$ is the Bose
distribution function at temperature $T$.

\begin{figure}[tbp]
\includegraphics*[width=0.5\linewidth, angle=90]{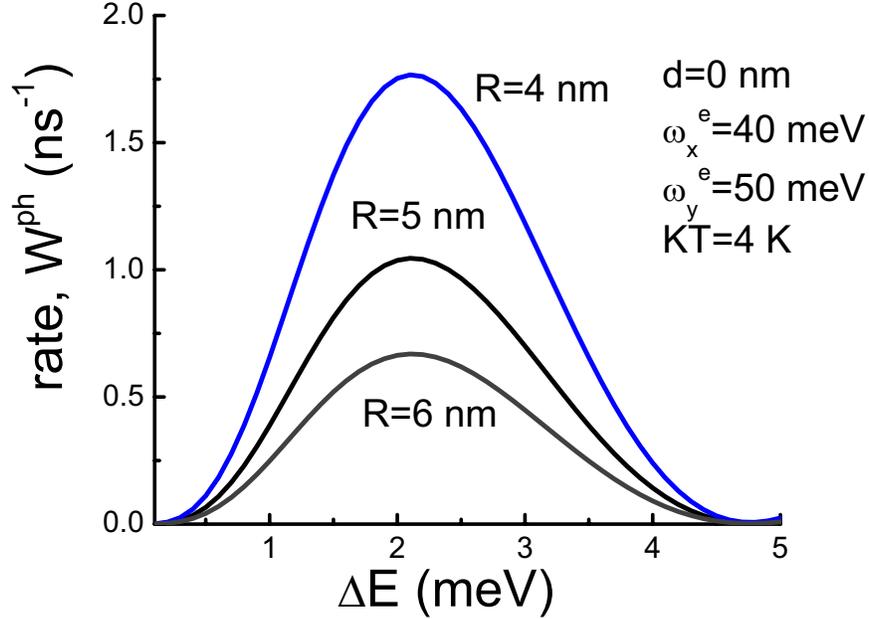}
\caption{ Calculated rates of phonon-assisted transfer as a
function of the resonance $\Delta E$ at $T=4~K$. The QD pair has
inversion symmetry, $d=0$.}  \label{fig6}
\end{figure}

The rates calculated from eq.~\ref{Wph2} strongly depend on the
energy difference $\Delta E$ (fig.~\ref{fig6}). At small $\Delta
E$ and low temperature $T$, the rate decreases due to the phonon
density of states, whereas at large $\Delta E$, it becomes small
due to the matrix elements of the function $e^{iqr}$. The
calculated rate is maximum at $\Delta E\sim2~meV$ and is about
$ns^{-1}$. Since the exciton-phonon interaction (\ref{Hph}) does
not include spin-dependent operators, the spin information is not
lost in the phonon-emission process. Therefore the spin-selection
rules are given by the Coulomb matrix elements while the phonon
matrix elements conserve the exciton spin configuration
\cite{comment}. In a symmetric pair of QDs with $d=0$, the spin of
exciton is conserved. However, it is important to note that we
have neglected the mixing between heavy and light holes; this
mixing together with the electron-phonon interaction can result in
an additional spin relaxation. In oblate QDs considered here, our
approximation is justified since the heavy hole-light hole mixing
is suppressed due to the large slitting between heavy- and
light-hole levels in the valence band.

\section{ Spin-dependent cross-correlation functions}

Similar to the molecular systems \cite{molecules}, the coupling
between QDs can be seen in the photon correlation measurements
\cite{Petroff-preprint}. Here we are going to introduce
spin-dependent correlation functions for the case of two coupled
QDs. The second-order correlation function is defined as
$g^{(2)}_{ij}=<I_i(t)I_j(t+\tau)>/<I_i(t)><I_j(t)>$, where
$I_i(t)$ is the emission intensity of the $i$-exciton state. The
function $g^{(2)}_{ij}$ is proportional to a number of photon
pairs arriving with time interval $\tau$.

The nonlinear dynamics of a double-dot system can be quite
complex. For simplicity, we will consider the limit of weak
pumping when the biexciton contribution to the density matrix is
small. Assuming nonresonant unpolarized excitation of low
intensity, we can describe the exciton dynamics with a system of
linear equations:

\begin{figure}[tbp]
\includegraphics*[width=0.7\linewidth, angle=90]{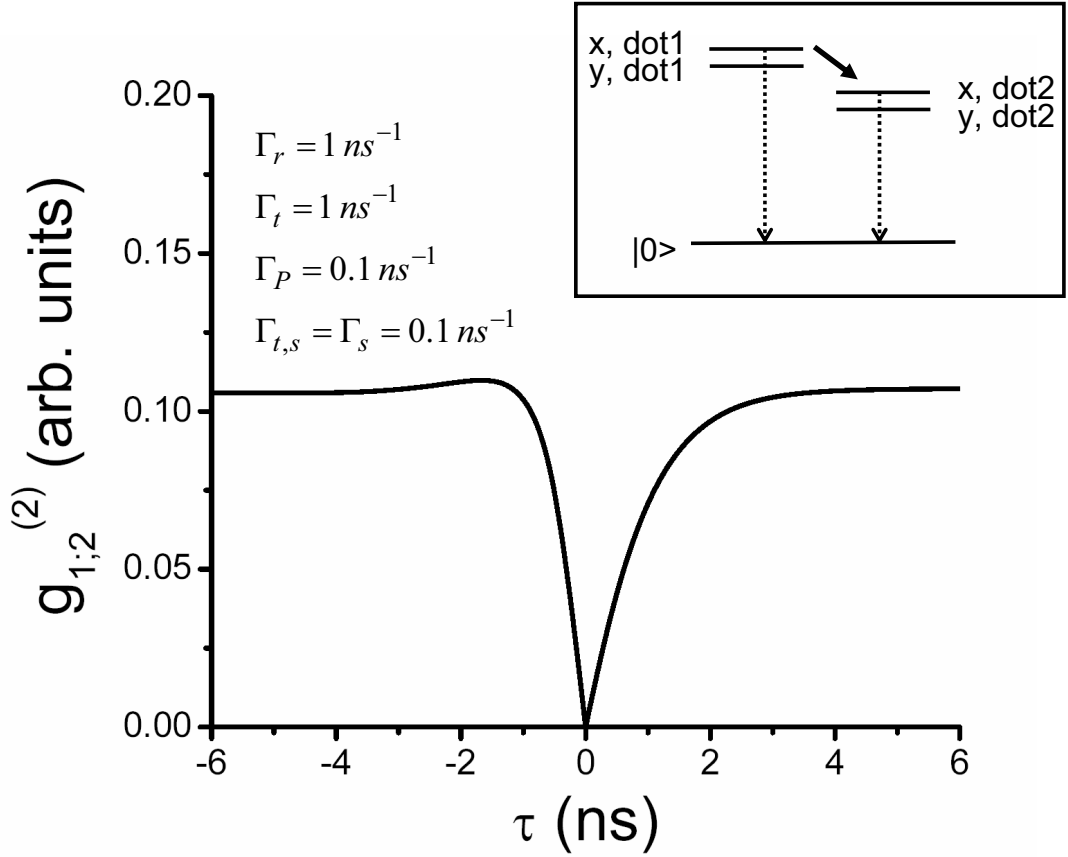}
\caption{Calculated cross-correlation functions for a QD pair with
one exciton. The correlation function is independent of spin state
of excitons. The parameters of relaxation are shown in the figure
and correspond to a QD pair with $R=5~nm$, $d=0$, and $\Delta
E\sim2~meV$. Insert: the energy diagram.} \label{fig7}
\end{figure}

\begin{eqnarray}
\nonumber
\dot{n}_{0}=\Gamma_r(n_{x,1}+n_{y,1}+n_{x,2}+n_{y,2})-4\Gamma_Pn_0,
\\ \nonumber
\dot{n}_{x,1}=-(\Gamma_r+\Gamma_t+\Gamma_{t,s}+\Gamma_s)n_{x,1}+\Gamma_sn_{y,1}+\Gamma_Pn_0,
\\ \nonumber
\dot{n}_{x,2}=-(\Gamma_r+\Gamma_s)n_{x,2}+\Gamma_{t,s}n_{y,1}+\Gamma_sn_{y,2}+\Gamma_Pn_0,\\
\nonumber
\dot{n}_{y,1}=-(\Gamma_r+\Gamma_t+\Gamma_{t,s}+\Gamma_s)n_{x,1}+\Gamma_sn_{x,1}+\Gamma_Pn_0,
\\
\dot{n}_{y,2}=-(\Gamma_r+\Gamma_s)n_{y,2}+\Gamma_{t,s}n_{x,1}+\Gamma_sn_{x,2}+\Gamma_Pn_0,
\label{dynamics}
\end{eqnarray}
where $n_0$ is the "vacuum" exciton state, $n_{\alpha,k}$ are the
numbers of excitons, $\alpha=x,y$ is the type of exciton at $B=0$,
and $k=1,2$ is the QD number. The rate $\Gamma_r$ describes
radiative recombination, $\Gamma_t$ is the energy transfer rate
from QD1 to QD2 with conservation of spin, $\Gamma_{t,s}$ is the
inter-dot transfer rate with spin flip, $\Gamma_s$ is the
intra-dot spin-flip rate, and $\Gamma_P$ is the pumping rate
proportional to the light intensity. This simple model resembles
the ones used in refs.~\cite{Petroff-preprint,Michler}. For our
calculations we choose: $\Gamma_t=1~ns^{-1}$,
$\Gamma_r=1~ns^{-1}$, $\Gamma_{t,s}=\Gamma_s=0.1~ns^{-1}$, and
$\Gamma_P=0.1~ns^{-1}$. The pumping rate $\Gamma_P=0.1~ns^{-1}$
corresponds to the regime of low intensity ($n_0\sim1$). The small
phenomenological spin-flip rates are chosen to take into account
spin-flip events which are typically slow. The calculated
cross-correlation functions are shown in
figs.~\ref{fig7},\ref{fig8}.

\begin{figure}[tbp]
\includegraphics*[width=0.7\linewidth, angle=90]{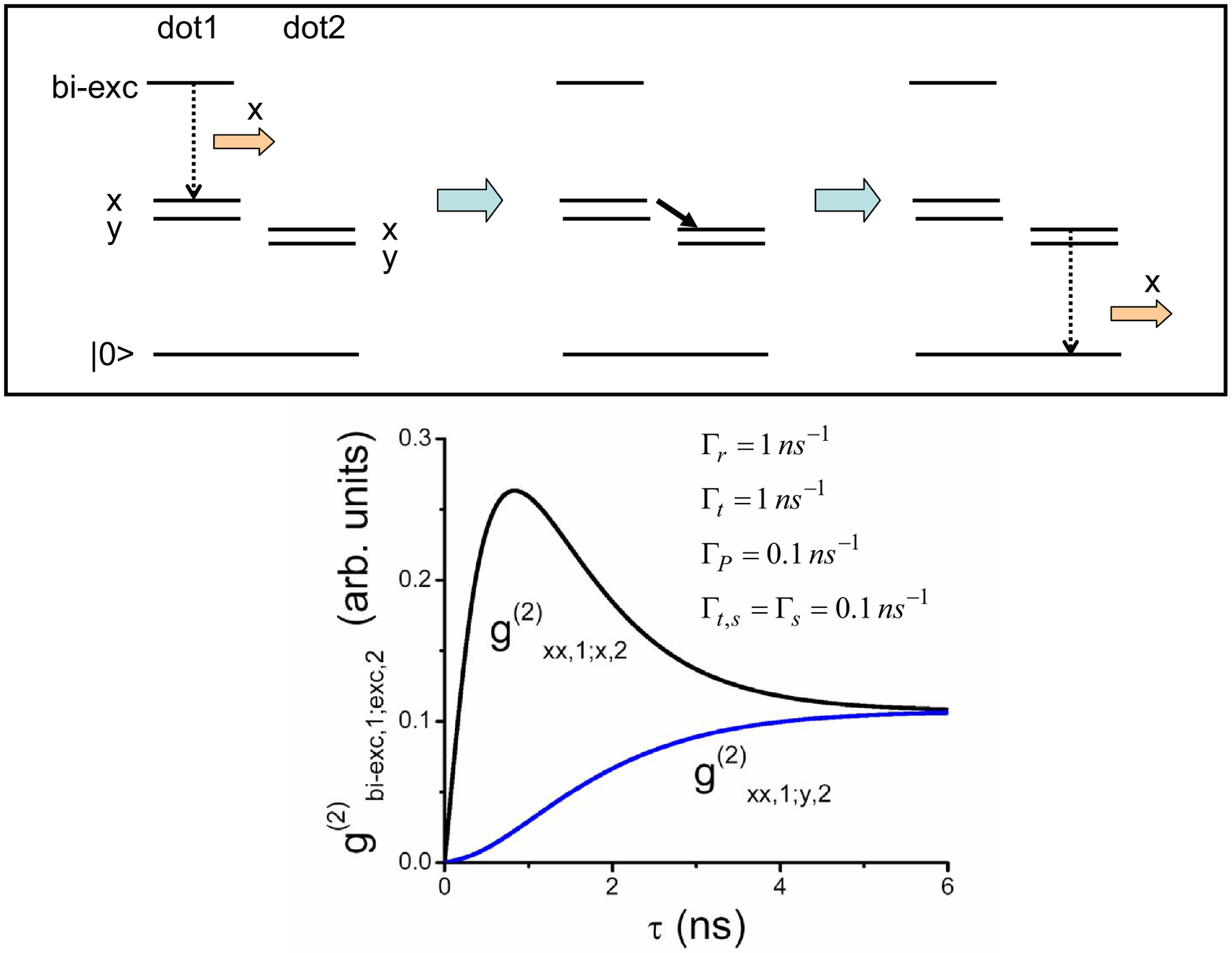}
\caption{ Upper part: schematics of processes contributing to the
correlation function $g^{(2)}_{xx,1;x,2}$. In the first step, a
bi-exciton in QD1 emits the $x$-photon; the second step is
inter-dot transfer; the last step is emission of the $x$-photon by
QD2. Lower part: calculated functions $g^{(2)}_{xx,1;x,2}$ and
$g^{(2)}_{xx,1;y,2}$. Since inter-dot transfer mostly conserves
spin,  $g^{(2)}_{xx,1;x,2} > g^{(2)}_{xx,1;y,2}$. The parameters
of relaxation are shown in the figure and correspond to a QD pair
with $R=5~nm$, $d=0$, and $\Delta E\sim2~meV$. } \label{fig8}
\end{figure}

First we describe one-exciton cross-correlation functions
$g^{(2)}_{x,1;x,2}$ and $g^{(2)}_{x,1;y,2}$ which turn out to be
spin-independent in our model for the weak pumping regime:
$g^{(2)}_{x,1;x,2}=g^{(2)}_{x,1;y,2}=g^{(2)}_{1;2}$. The time
delay $\tau$ is positive when the photon with energy
$E_{exc,dot1}$ arrives before the photon with $E_{exc,dot2}$. At
$\tau=0$, emission of the photon $E_{exc,dot1}$ projets the system
from the state $|dot1,x>$ to the "vacuum" state. Therefore, the
initial conditions for eqs.~\ref{dynamics} are set as: $n_0=1$ and
$n_{\alpha,k}=0$. For $\tau>0$, $g^{(2)}_{1;2}\propto
n_{x,2}(\tau)=n_{y,2}(\tau)$, where $n_{x(y),2}(\tau)$ are the
solutions of eqs.~\ref{dynamics} for the above initial conditions.
For $\tau<0$, $g^{(2)}_{1;2}\propto n_{x,1}(\tau)=n_{y,1}(\tau)$.
The function $g^{(2)}_{1;2}$ is not symmetric with respect to
$\tau$ because of directional exciton transfer from QD1 toward
QD2. The effective exciton lifetime in QD1 is shorter since an
exciton can be transferred to QD2. This is reflected as a faster
increase of $g^{(2)}_{1;2}$ at $\tau<0$. In a magnetic field, the
cross-correlation functions can become polarization-dependent
($g^{(2)}_{x,1;x,2}\neq g^{(2)}_{x,1;y,2})$ since the resonance
conditions can be different for various transfer processes. This
can be incorporated in the model through appropriate
spin-dependent transfer rates $\Gamma_{t,J_{tot}\rightarrow
J_{tot}'}$.

Spin transfer processes can be observed using the bi-exciton
lines. Even at small pumping, weak bi-exciton lines exist. The
energy of bi-exciton lines are red-shifted by few meV and can be
distinguished from the one-exciton lines. In addition, the
bi-exciton lines have a quadratic power dependence. Consider now
the bi-exciton in QD1. It decays in a radiative cascade emitting
two photons with the same polarizations ($x$, $x$ or $y$, $y$
photon pairs) \cite{Michler2}. If the first emitted photon has the
$x$-polarization, the remaining exciton in QD1 has the
$x$-character. This exciton can recombine or can be transferred to
QD2. Since the exciton spin is mostly conserved in the transfer
process, we expect that QD2 will strongly radiate $x$-photons
shortly after emission of the $x$-photon from  QD1. This means
that $g^{(2)}_{xx,1;x,2} > g^{(2)}_{xx,1;y,2}$, where the index
$xx$ labels the $x$-photon emitted by the bi-exciton in QD1. In
other words, we use a bi-exciton in QD1 as a tool to prepare the
$x$-exciton state at $\tau=0$. Then, the initial conditions at
$\tau=0$ are: $n_0=0$, $n_{\alpha,k}=1$ if $(\alpha,k)=(x,1)$ and
$0$ otherwise. Figure~\ref{fig8} demonstrates the striking
difference between the polarized correlation functions
($g^{(2)}_{xx,1;x,2}$ and $g^{(2)}_{xx,1;y,2}$) in the important
region $\tau>0$. In this way, by comparing $g^{(2)}_{xx,1;x,2}$
and $g^{(2)}_{xx,1;y,2}$, directional spin transfer between QDs
can be observed experimentally.

\section{Strongly-resonant Coulomb transfer}

\begin{figure}[tbp]
\includegraphics*[width=0.4\linewidth, angle=90]{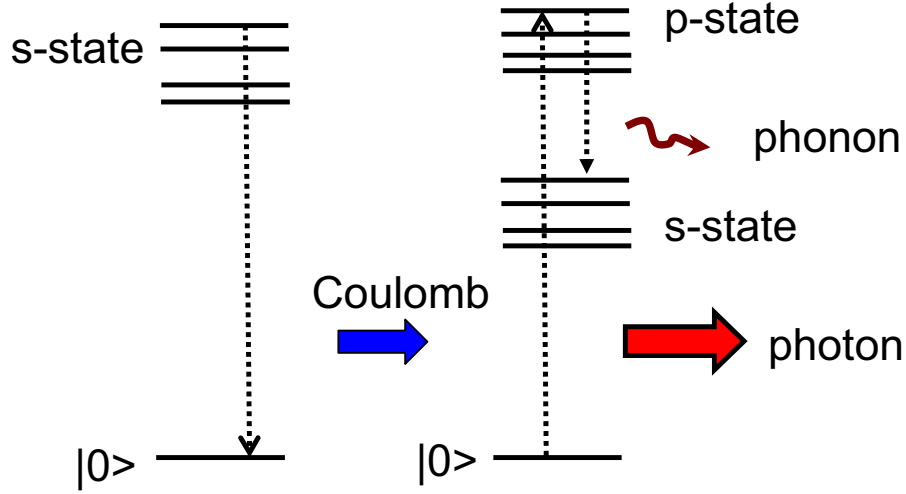}
\caption{ Resonant transfer process involving the s-state in QD1
and the p-state in  QD2. Relaxation in  QD2 is phonon-assisted.}
\label{fig9}
\end{figure}

The convectional F\"orster mechanism is based on the resonance
condition between the "donor" and "acceptor" \cite{Forster}: the
ground-state energy of the donor molecule coincides with the
energy of an excited state of the acceptor. In the QD system, such
a condition can be realized if the ground $s-s$ exciton transition
in the QD1 has the same energy as the $p-p$ transition in QD2
(fig.~\ref{fig9}). Here $s$ and $p$ are the shell indexes in a QD.
In this process, the $s$-exciton in QD1  is first transferred to
the p-state of QD2; then it relaxes to the ground state of the
QD2. The transfer rate of this process consists of two
contributions:

\begin{eqnarray}
W^{res}_{dot1\rightarrow dot2}=W^{ph}+W^{dir}, \label{Wres}
\end{eqnarray}
where $W^{ph}$ is the phonon-assisted transfer rate given by
eq.~\ref{Wph} in which $\Delta E$ is the energy difference between
$s$- and $p$-excitons in QD1 and QD2, respectively. The rate
$W^{dir}$ describes direct resonant transfer between QDs. The
latter can be calculated in the spirit of the convectional
F\"orster theory as

\begin{eqnarray}
W^{dir}=\frac{2\pi}{\hbar}|M_{dot1,\alpha\rightarrow
dot2,\alpha'}|^2J(\Delta E), \label{Wres1}
\end{eqnarray}
where $J(\Delta E)=\frac{1}{\hbar\pi}\frac{\Gamma_{en}/2}{\Delta
E^2/\hbar^2+\Gamma_{en}^2/4}$ is the normalized effective density
of states in the QD2, $\Gamma_{en}$ is the energy relaxation rate
in the QD2, and $\Delta E=E_{exc,dot1}-E_{exc,dot2}$. To obtain
the equation (\ref{Wres1}) one should solve the master equation
involving the density matrix and assume that
$\hbar\Gamma_{en}>|M_{dot1\rightarrow dot2}|$. The latter
condition can be easily satisfied because the excited p-states are
quasi-stationary and the phonon-induced relaxation in QDs is fast
usually; the typical relaxation times of self-assembled QDs are in
the range of $50~ps$ \cite{comment2}. In the opposite limit
$\hbar\Gamma_{en}<|M_{dot1\rightarrow dot2}|$, the coupling
between the ground $s$-state of QD1 and the excited $p$-state of
QD2 is coherent; it has the character of Rabi oscillations. Then,
in the case of $\hbar\Gamma_{en}<|M_{dot1\rightarrow dot2}|$, the
characteristic time for transfer from the $s$-state of QD1 to the
$s$-state of QD2 will be about $1/\Gamma_{en}$.

\begin{figure}[tbp]
\includegraphics*[width=0.5\linewidth, angle=90]{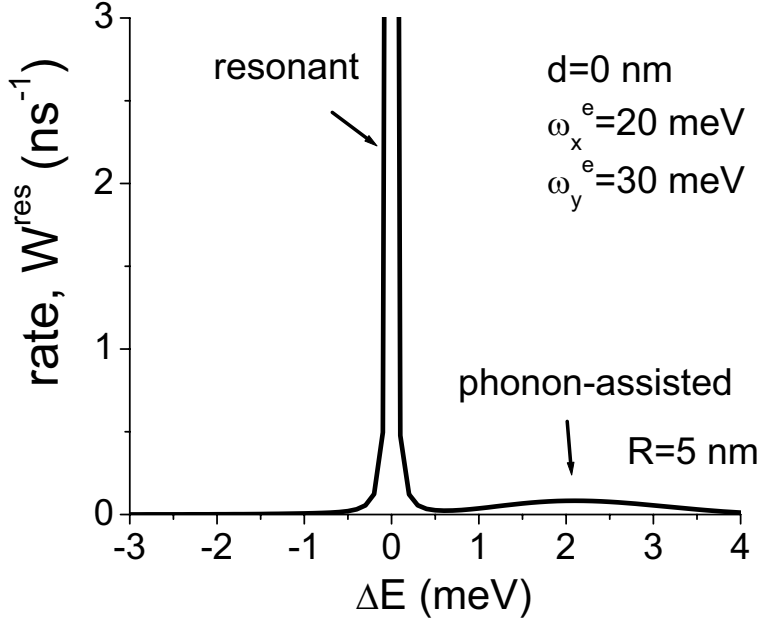}
\caption{ Resonant transfer rate corresponding to the process
shown in fig.~8. The parameters of the QD pair are: $R=5$,
$\hbar\omega^e_x=20~meV$, $\hbar\omega^e_y=30~meV$, $d=0$, and
$1/\Gamma_{en}=40~ps$. Temperature $T=4~K$. The exciton spin is
conserved in the transfer process. } \label{fig10}
\end{figure}

We now calculate the resonant transfer rate for typical parameters
of self-assembled QDs. Again, the rate strongly depends on the
resonance condition (see fig.~\ref{fig10}). The calculated rate
demonstrates strong enhancement for $\Delta E\sim\hbar\Gamma_{en}$
and also has the structure due to the phonon-assisted processes
discussed above.

The transfer rate $W^{res}$ is proportional to
$|M_{dot1,\alpha\rightarrow dot2,\alpha'}|^2$ and therefore the
spin-selection rules are given by the Coulomb matrix elements.
However, the complete transfer process contains energy relaxation
inside QD2 (fig.~\ref{fig9}). This relaxation can lead to spin
flip. Then, the efficiency of spin transfer will also depend on
the ratio between intra-dot relaxation rates with and without spin
flip. If energy relaxation inside QD2 involves mostly the
heavy-hole states, spin-flip relaxation will be weak, since the
main contributions to electron-phonon scattering are not
spin-dependent and, simultaneously, the heavy-hole exciton
functions are factorized \cite{comment}. It concerns both the
acoustic-phonon interaction (\ref{Hph}) and the Fr\"ohlich
scattering with emission of LO-phonons \cite{Maham}. To conclude
this section,  we note that the resonant transfer process can
involve a localized state in the wetting layer, instead of the
$p$-state in QD2. Such a possibility was discussed in
ref.~\cite{Petroff-preprint}.

\section{Discussion}

In sections 3 and 5, we discussed incoherent transfer assisted by
phonons or involving a broadened state in  QD2. The
phonon-assisted transfer regime between ground states in QD1 and
QD2 assumes that the energy difference $\Delta E$ is larger than
the Coulomb matrix element $|M_{dot1\rightarrow dot2}|$. Now we
briefly consider the case of coherent resonant coupling in the
regime $\Delta E\sim|M_{dot1\rightarrow dot2}|$
\cite{coherent,coherent2,coherent3}. This regime requires fine
tuning of energies of QDs; this can be done, for example, with
magnetic and electric fields. The QDs can be designed from
different materials and therefore may have different $g$-factors.
By changing a magnetic field, one can change $\Delta E$. A similar
principle can be used in the case of an applied electric field; if
the QDs have different dipole moments, $\Delta E$ can be
controlled by the electric field.

The calculated Coulomb matrix elements (see fig.~\ref{fig4}) are
in the range of $0.05~meV$ for $R=5~nm$. The corresponding time is
quite short: $\Delta t=\hbar/|M_{dot1\rightarrow dot2}|\sim10~ps$.
This time is much shorter than the typical recombination time of
ground-state excitons in QDs, which is about $1~ns$. Therefore the
coherent Coulomb-induced coupling can exist in resonant pairs of
QDs. In the case of $\Delta E=0$ and a QD molecule with $d=0$, the
one-exciton wave functions are given by the linear combinations:

\begin{eqnarray}
\Psi_x=\frac{|dot1,x>|dot2,0>\pm|dot1,0>|dot2,x>}{\sqrt{2}},
\nonumber  \\
\Psi_y=\frac{|dot1,y>|dot2,0>\pm|dot1,0>|dot2,y>}{\sqrt{2}}.
\label{PsiRes}
\end{eqnarray}
The energy splitting within the pairs of states is given by
$2|M_{dot1\rightarrow dot2}|$. This energy can be regarded as a
Rabi frequency. If the exciton is created initially in the QD1,
the time to transfer the exciton to the neighboring dot would be
$\Delta t=\hbar/|M_{dot1\rightarrow dot2}|\sim10~ps$ for the QD
pair with $R=5~nm$. With increasing $R$, the transfer times will
become longer. In the regime of coherent coupling, transfer of
spin information occurs coherently, without any energy
dissipation. The coherent spin-Rabi oscillations between QDs can
probably be observed with modern optical methods.

To conclude, we have described the spin-dependent Coulomb
interaction in a QD pair. Such a coupling suggests the possibility
to transfer spin information between individual nano-crystals
without transfer of charge or mass. The spin-dependent transfer
originates from the exchange and spin-orbit interactions in
semiconductors and strongly depends on symmetry and shapes of
nano-crystals. If symmetry of a QD pair is high enough, spin
information can be transferred without losses. If symmetry is
broken, spin relaxation in the transfer process can become
significant. To calculate the transfer rates in the realistic
model, we use a generalized dipole-dipole approximation which is
valid if $R\gg a_{lattice}$. The usual dipole-dipole approximation
($R\gg l_{dot}$) gives too large numbers for the realistic
inter-dot distances. As a method to observe spin transfer, we
consider spin-dependent photon correlations in a pair of QDs.

{\bf Acknowledgements}. The author would like to thank Pierre
Petroff for some important comments and S.~Ulloa, G.~Bryant,
B.~McCombe, G.~Medeiros-Ribeiro, M.~Ouyang, and M.~Bayer for
helpful discussions. This work was supported by the CMSS Program
at Ohio University and the AvH and Volkswagen Foundations.


\begin{thebibliography}{99}
\bibitem{commentHL} Our model does not include the heavy-light hole
mixing which may involve dark states in transfer processes.
\bibitem{comment} It is also important to note that the heavy-hole exciton states
can be written as a product of the envelope function and the
spin-dependent Bloch function, $|dot \
k>=\phi_e(r_e)\phi_h(r_h)\chi_{exc}(r_e,\sigma_e,r_h,\sigma_h)$,
where $\chi_{exc}$ is the Bloch function of an exciton. The
electron-phonon operator in the matrix elements act only on the
spin-independent envelope function $\phi_e(r_e)\phi_h(r_h)$ and
therefore the phonon matrix elements are diagonal with respect to
the exciton spin index $\alpha$.
\bibitem{comment2}
This time is of order of the time to trap an electron-hole pair
into the ground state of a QD and is typically short.
\end{thebibliography}
\end{document}